\documentclass[twocolumn]{revtex4}

\usepackage{amsmath}

\usepackage{graphicx}

\newcommand{\refe}[1]{\unskip~(\ref{#1})}
\newcommand{\av}[1]{\langle #1 \rangle}

\newcommand{\be}{\begin{equation}}
\newcommand{\bel}[1]{\begin{equation}\label{#1}}
\newcommand{\ee}{\end{equation}}
\newcommand{\bea}{\begin{eqnarray}}
\newcommand{\ea}{\end{eqnarray}}

\newcommand{\Erel}{E_\text{rel}}
\newcommand{\Ecm}{E_\text{cm}}
\renewcommand{\vec}[1]{\mathbf{#1}}

\newcommand{\sbsqrt}[1]{\sqrt{\smash[b]{#1}}} 

\begin{document}

\title{Averaged collision and reaction rates in a two-species gas of ultracold fermions}
\begin{abstract}
Reactive or elastic two-body collisions in an ultracold gas are affected by quantum statistics.
In this paper, we study ensemble-averaged collision rates for a two-species fermionic gas. 
The two species may have different masses, densities and temperatures. 
We investigate how averaged collision rates are affected by
the presence of Fermi spheres in the initial states 
(Pauli blocking of final states is not considered).
It is shown that, independently on the details of the collision,
Fermi-averaged collision rates deviate from Boltzmann-averaged ones,
particularly for a gas with strong imbalance of masses or densities.
\end{abstract}  
\author{Alexander Pikovski}
\affiliation{\mbox{$^{}$Institut f\"ur Theoretische Physik, Leibniz Universit\"at Hannover, Appelstr.~2, 30167 Hannover, Germany}}

\maketitle

\section{Introduction}

Experiments with ultracold atoms and molecules probe collisional physics
at very low temperatures. 
Gases of quantum degenerate fermionic atoms are being studied~\cite{Ketterle2008}, 
and experiments are progressing towards the creation of degenerate gases of fermionic molecules~\cite{Carr2009}. 
%
At very low temperatures, quantum statistics plays an important role in 
interparticle collisions.
%
One way to observe an effect which depends on the symmetry of the particles
is to look at low-energy collisions 
which obey quantum threshold laws;
this was measured in elastic collisions in fermionic atomic gases \cite{DeMarco1999}
and in reactive collisions of ultracold molecules \cite{Ospelkaus2010}.
Another effect of quantum statistics is Pauli blocking of collisions,
as was observed in Ref.~\cite{DeMarco2001}:
collisions are suppressed if the final state cannot enter a filled Fermi sphere.
In contrast to the quantum threshold behavior, the observation of this effect requires
the fermions to be quantum degenerate. 

In the present work we investigate how averaged collisional rates depend on the temperature,
density, and masses in a two-species Fermi gas.
The usual experimental observable in a cloud of ultracold gas is a
collision rate which is averaged over the distribution of velocities in the gas.
Usually this ensemble-averaging of collision rates is done using a Boltzmann distribution,
however as experiments move deeper into the quantum degenerate regime, 
quantum statistics starts playing a role. 
Particular attention here will be focused on the imbalanced Fermi gas, 
i.e.~a two-component gas where the two species have different densities or masses. 
The effect of mass or density imbalance on the averaged collision
rates at low temperatures is most pronounced in this setting, the rates can differ 
significantly from a Boltzmann-averaged rate.

Averaged collision rates for a two-component gas of fermions are studied, for
general two-body collisions with an arbitrary cross-section.
The particles are assumed to be in equilibrium; effects of Pauli-blocking are not
considered. 
The two types of fermions may have different masses, different densities, and different temperatures.
First, it is shown that an averaged collision rate is expressed as an 
average over the distribution of relative energies
(Sec.~\ref{sec-rate}).
This distribution is calculated for fermions analytically at $T=0$,
and it is shown that this function has a simple geometric interpretation
(Sec.~\ref{sec-DRE}). 
At finite temperatures, the relative energy distribution is calculated numerically
and a simple approximation for all temperatures is proposed 
(Sec.~\ref{sec-DRE}). 
In Sec.~\ref{sec-thres} we discuss the threshold laws for ensemble-averaged
collision rates. Some details of the calculations are presented in the Appendices.
In Sec.~\ref{sec-2D}, we discuss how the results can be transferred to a situation
where the particles are confined to a two-dimensional geometry.


\section{Averaged collision rate}
\label{sec-rate}

Experiments with clouds of ultracold gases measure collision rates,
e.g.~using photoassociation spectra~\cite{Napolitano1994}
or trap loss rates due to chemical reactions~\cite{DeMiranda2011}.
Here one actually measures rates which are averaged over the distribution of velocities
in the gas.
This ensemble-averaging is the topic of this section.

Consider a collision between two particles, $A$ and $B$, which results in some final state $f$:
\begin{equation}\label{proc}
A + B \to f . 
\end{equation}
This may be an elastic collision, an inelastic collision, or a reaction.
The particles $A$, $B$ have masses $m_A$, $m_B$ and velocities $\vec{v}_A$, $\vec{v}_B$
in the laboratory frame. 
The total scattering cross-section for the process \refe{proc} in the center-of-mass frame will be denoted $\sigma(p)$,
it depends on the relative momentum $p$ of the colliding particles.

In a gas cloud, where many collisions take place, the number density $n_\alpha$ of particles 
in state $\alpha=f,A,B$ satisfies the rate equation
\begin{equation}\label{rateeq}
 \frac{d n_f}{dt} = \av{v \sigma(p)} \, n_A n_B 
\end{equation} 
where $\av{v \sigma}$ is the two-particle collision rate averaged over the distribution
of velocities in the system. The rate equation \refe{rateeq} refers only to the process \refe{proc}.
It is assumed that, in the case of fermions, the final state $f$
is not occupied, thus there is no effect of Pauli blocking.
If the particles $A$ and $B$ are indistinguishable, a small modification to Eq.~\refe{rateeq} may be necessary,
see~\cite{Julienne1989}.

The relevant quantity for many experiments is the ensemble-averaged collision rate
$K=\av{v \sigma}$, it has dimensions $(\text{length})^3/\text{time}$.
If velocities of particles of type $A$, $B$ are distributed according to the distributions
$\bar{f}_A$, $\bar{f}_B$, the averaged collision rate $K$
is \cite{Shuler1968}
\begin{equation}\label{K3d}
K = \langle v \sigma(p) \rangle = \iint v \sigma(p) \bar{f}_A(v_A) \bar{f}_B(v_B) \, d^3v_A d^3v_B
\end{equation} 
We will consider the case where the velocity distributions and the cross-section 
depend only on the magnitude of the velocities, 
i.e.~they depend only on the energy of the particles.
Changing to the center-of-mass frame with the formulas of Appendix~\ref{app-transf}
brings the integral to the following form:
\begin{equation}\label{K-F3d}
K = \sqrt{\frac{2}{m}} \, \int_0^\infty \!\! \Erel \sigma(\Erel) F(\Erel) \, d\Erel
\end{equation}
with
\begin{equation}\label{F3d}
F(\Erel) \!=\! \frac{1}{2} \int_0^\pi \!\!\! d\theta \sin\theta \!\! \int_0^\infty \!\!\!  f_A(E_A) f_B(E_B) 
\Ecm^{1/2}
d\Ecm .
\end{equation} 
Here $f_{A,B}$ are the energy distribution functions.
The normalization for the velocity distribution functions is $\int \! \bar{f}(v) d^3v=1$,
the energy distribution functions are normalized to $\int_0^\infty \! f(E) \sqrt{E} dE=1$,
and $f_{A,B}= 2\pi (2 / m_{A,B})^{3/2} \bar{f}_{A,B}$.
The expressions for $E_A$ and $E_B$ in terms of $E_\text{cm}$, $E_\text{rel}$ and $\theta$, which are needed
to evaluate the integral~\refe{F3d}, are given in Eq.~\refe{EAB-CR}.

The ensemble-averaged collision rate is given by the cross-section $\sigma$
averaged with the function $F$, as shown by Eqs.~\refe{K-F3d}, \refe{F3d}.
The function $F(\Erel)$, which we call distribution of relative energies,
contains all the dependence on the densities and temperatures of the particles
involved in the collision. The cross-section $\sigma(p)$ contains
all the information about the two-body scattering.

The expression for the distribution function of relative energies,
Eq.~\refe{F3d}, can be brought to a different form which will be of use later.
Transforming the integral in Eq.~\refe{F3d} to the coordinate system $(x,y,z)$, as described in
Appendix~\ref{app-transf}, we have the representation
\begin{equation}\label{F-xyz}
 F(\Erel) = \frac{1}{2\pi \eta^3} \iiint f_A(E_A) f_B(E_B) \, dxdydz .
\end{equation}  
The integration runs over the whole space, $E_A$ and $E_B$ are expressed
through $x, y, z$ using Eqs.~\refe{EAB-CR}, \refe{Exyz}, and the mass factor $\eta$ is defined in Eq.~\refe{eta}.

\section{Distribution of relative energies for fermions}
\label{sec-DRE}

Now we discuss the distribution function of relative
energies $F(\Erel)$, which determines the averaged collision rate [cf.~Eq. \refe{K-F3d}]
for a two-species gas of fermions in equilibrium.
We will consider a Fermi gas where the two species may have different densities or masses. 
Such imbalanced systems have attracted recent experimental \cite{Partridge-Shin}
and theoretical \cite{Chevy-Radzhihovsky} attention.

\subsection{Zero temperature}

\begin{figure}[t]
\centering
\includegraphics[width=0.6\columnwidth]{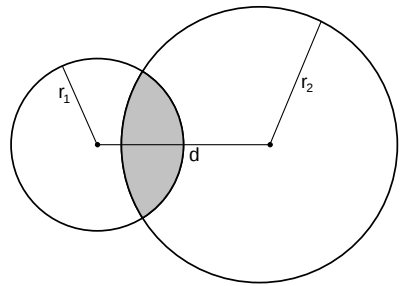}
\caption{Intersection of two spheres with radii $r_1$ and $r_2$, centers separated by distance $d$.
}\label{fig:spheres}
\end{figure}%

First, the case of zero temperature is considered.
The distribution functions for the particles $A$ and $B$
are normalized $T=0$ Fermi functions
\begin{equation}\label{f0}
f_{0}(E)=\frac{3}{2\mu^{3/2}} \; \theta(\mu-E),
\end{equation} 
where $\theta(x)$ is the step function and $\mu$ is the chemical potential.
The particles of type $A$, $B$ may have different chemical potentials $\mu_A$, $\mu_B$.
It will be convenient to introduce the ``reduced'' chemical potentials
\begin{equation}\label{nu}
\nu_A = \frac{2 m_B}{m_A+m_B} \mu_A,\qquad
\nu_B = \frac{2 m_A}{m_A+m_B} \mu_B ,
\end{equation} 
and for equal masses we have $\nu_{A,B}=\mu_{A,B}$.
When the Fermi functions, Eq.~\refe{f0}, are inserted in the representation \refe{F-xyz} for
the distribution of relative energies, one encounters the volume integral
\begin{equation}\label{V-int}
\int \theta(\mu_A - E_A ) \theta(\mu_B - E_B ) d^3x .
\end{equation} 
Once $E_A$ and $E_B$ are expressed in terms of $x,y,z$, it seen that this integral has the following geometric
meaning: it is the volume of intersection of two spheres. 
Let us briefly discuss this volume.

The volume of intersection of two spheres, which have radii $r_1$ and $r_2$ and whose centers are separated
by a distance $d$ (see Fig.~\ref{fig:spheres}), is given by the expression
\begin{equation}
 V_L(d;r_1,r_2) \!=\! \frac{\pi}{12d} (r_1 + r_2 - d)^2 \{ d^2 + 2d(r_1+r_2) - 3(r_1-r_2)^2 \} .
\end{equation} 
This formula, as it stands, is only valid as long as the volume of intersection has the shape of an asymmetric
lens, as indicated in Fig.~\ref{fig:spheres}. However, if the distance between the spheres $d$ is large, there is no intersection, 
and if $d$ is small, one sphere is located completely inside of the other. 
We can write for the volume of intersection
\begin{equation}\label{V}
V(d;r_1,r_2) =
\left\{
\begin{array}{ll}
0 & \quad\text{if } d^2 > (r_1+r_2)^2 \\[2pt]
\min(\frac{4}{3}\pi r_1^3,\frac{4}{3}\pi r_2^3)  & \quad\text{if } d^2 < (r_1 - r_2)^2 \\[3pt]
V_L(d;r_1,r_2) & \quad\text{otherwise}
\end{array}
\right.
\end{equation} 
In this form the expression is valid for all values of $d$, $r_1$, and $r_2$.

Returning to the distribution function of relative energies,
it is seen by using~\refe{f0} in Eq.~\refe{F-xyz} that the result is
\begin{equation}\label{F-T0}
 F_0(\Erel; \nu_A, \nu_B) 
\!=\! \frac{9}{2\sqrt{2}\pi} \frac{1}{(\nu_A \nu_b)^{3/2}} \, V(\sbsqrt{2 \Erel}; \sbsqrt{\nu_A}, \sbsqrt{\nu_B}) .
\end{equation} 
This is the volume of intersection of two spheres with radii $\sbsqrt{\nu_A}$ and $\sbsqrt{\nu_B}$,
the centers of the spheres being separated by $\sqrt{2\Erel}$, up to normalization.
It is interesting to note that the masses of the particles do not appear explicitly, but enter only
through the reduced chemical potentials $\nu_A$, $\nu_B$.

\begin{figure}[t]
\centering
\raisebox{-2ex}{a)\;}
\raisebox{-\height}{\includegraphics[width=0.78\columnwidth]{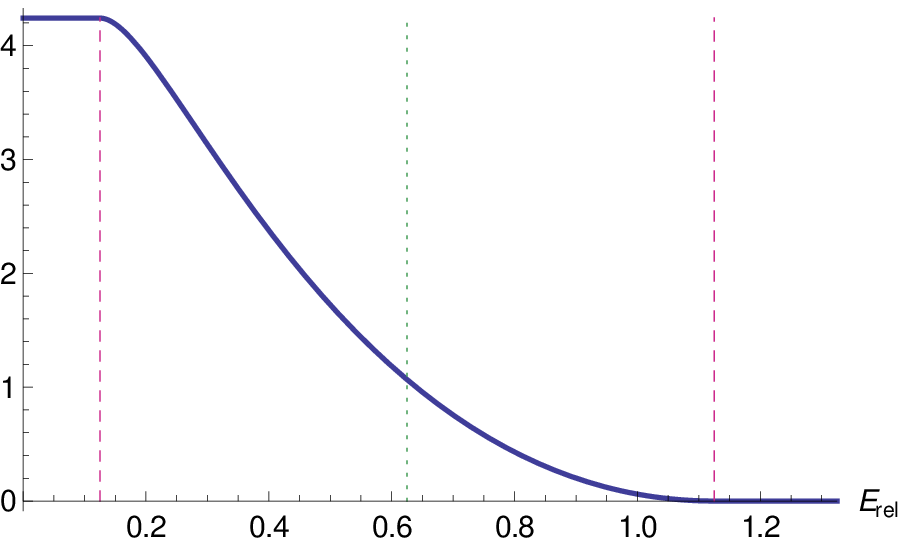}}\\
\raisebox{-2ex}{b)\;}
\raisebox{-\height}{\includegraphics[width=0.78\columnwidth]{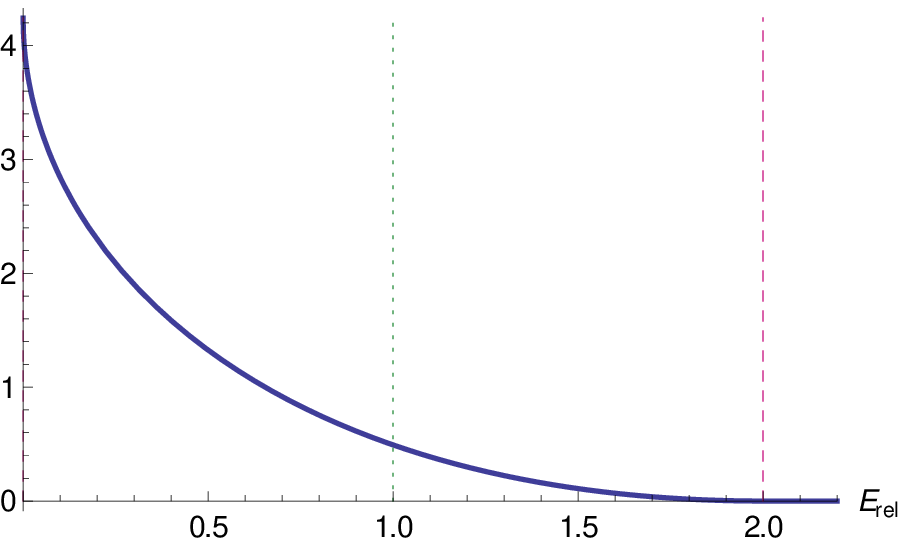}}
\caption{$F(\Erel)$ for a) $\nu_A=0.25$, $\nu_B=1.0$ and b) $\nu_A=\nu_B=1$. The dashed lines are at
$\frac{\nu_A+\nu_B}{2} - \sbsqrt{\nu_A \nu_B}$ and
$\frac{\nu_A+\nu_B}{2} + \sbsqrt{\nu_A \nu_B}$, the dotted line is at $\mu_\ast=\tfrac{\nu_A+\nu_B}{2}$.
}\label{fig:Fgen}
\end{figure}%

Let us discuss the general form of the function $F_0(\Erel)$.
A plot is shown in Fig. \ref{fig:Fgen}a for the strongly imbalanced case $\nu_1 \ll \nu_2$.
This case occurs either if the particles have different masses or if the particle 
densities are different, cf.~Eq.~\refe{nu}. The function has the form of a smoothed step.
From the cases in Eq.~\refe{V}, it is seen that $F_0(\Erel)$ is a decreasing function in the range
(shown by dashed lines in Fig.~\ref{fig:Fgen})
\begin{equation}
\frac{\nu_A + \nu_B}{2} - \sbsqrt{\nu_A \nu_B}  < \Erel < \frac{\nu_A + \nu_B}{2} + \sbsqrt{\nu_A \nu_B} .
\end{equation} 
For smaller values of $\Erel$ the function is constant and for larger values of $\Erel$ it is 
identically zero.
The step is centered at $\mu_\ast = (\nu_A +\nu_B)/2$, but it is not symmetric around this point.

In the balanced case $\mu=\nu_A=\nu_B$, the expression for $F_0(\Erel)$ simplifies considerably.
This case is shown in Fig.~\ref{fig:Fgen}b. 
The distribution function has no flat region, it is now a strictly decreasing function between $\Erel=0$ and $\Erel=2\mu$.

\begin{figure}[t]
\centering
\includegraphics[width=1.0\columnwidth]{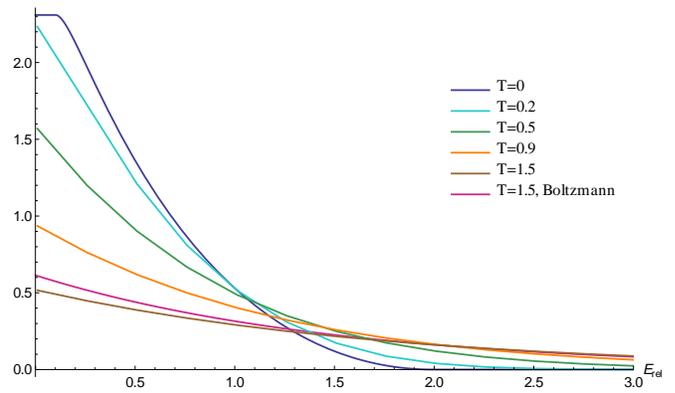}
\caption{$F(\Erel)$ at different temperatures, for $\nu_A=0.6$, $\nu_B=1.5$ at $T=0$.
}\label{fig:FgenT}
\end{figure}%

\subsection{Finite temperatures.}
For finite temperatures, the distribution functions of the particles are normalized Fermi--Dirac 
distributions
\begin{equation}
 f_T(E) =  \mathcal{N}_{\mu,T} \, \frac{1}{1+e^{(E-\mu)/T}},
\end{equation} 
the normalization constant ${\mathcal N}$ is given in Appendix \ref{app-finiteT}.
We allow the particles $A$, $B$ to be at different temperatures $T_A$, $T_B$.
The corresponding distribution of relative energies $F_T(\Erel)$ becomes, geometrically, 
the volume of overlap of two spherical distributions whose radii are broadened by temperature.
One can introduce the reduced temperatures
\begin{equation}\label{tau}
\tau_A = \frac{2 m_B}{m_A+m_B} T_A,\qquad
\tau_B = \frac{2 m_A}{m_A+m_B} T_B  .
\end{equation} 
As shown in Appendix \ref{app-finiteT}, the distribution of relative energies 
can be written as 
\begin{multline}\label{F-T}
 F_T(\Erel) = 
\frac{ {\mathcal N}_{\nu_A,\tau_A} {\mathcal N}_{\nu_B,\tau_B} }{ 36 \, \tau_A \tau_B} 
\\[4pt] \times
\int_0^\infty \!\!\! \int_0^\infty \!\!\! 
\frac{(\epsilon_A \epsilon_B)^{3/2} F_0(\Erel; \epsilon_A, \epsilon_B) }
{\cosh^2 \bigl( \frac{\epsilon_A-\nu_A}{2\tau_A} \bigr) \cosh^2 \bigl( \frac{\epsilon_B-\nu_B}{2\tau_B} \bigr)}
d\epsilon_A d\epsilon_B .
\end{multline}
Note that the masses of the particles enter only through the reduced chemical potentials
$\nu_\alpha$ and temperatures $\tau_\alpha$.

In the limit of high temperatures, both particles are distributed according to the Boltzmann distribution
\begin{equation}
f_\infty(E)=\frac{2}{\sqrt{\pi} T^{3/2}} \; e^{-E/T} .
\end{equation} 
In this case the distribution of relative energies may be evaluated directly from Eq.~\refe{F3d}.
Then we obtain, as known~\cite{Shuler1968},
that the distribution of relative energies is a Boltzmann distribution 
\begin{equation}\label{Boltz-3D}
F_\infty(E_\text{rel}) = \frac{2}{\sqrt{\pi} T^{3/2}_\ast} \; e^{-E_\text{rel} / T_\ast},  
\end{equation} 
with the effective temperature
\begin{equation}\label{Tast}
T_\ast = \frac{\tau_A + \tau_B}{2} =  \frac{m_A T_B + m_B T_A}{m_A+m_B} .
\end{equation} 

The distribution function of relative energies
may be calculated numerically from Eq. \refe{F3d} or from Eq. \refe{F-T}; for low temperatures 
the form in Eq.~\refe{F-T} is more convenient.
Figure~\ref{fig:FgenT} shows the result of such a calculation.
The chemical potentials $\nu_A$, $\nu_B$ were fixed at $T=0$, and at $T>0$ the
temperature-dependent chemical potentials were found by solving Eq.~\refe{mu-T}.
The figure also compares the distribution function with the Boltzmann
[$F_\infty(\Erel)$ from Eq.~\refe{Boltz-3D}] limit.
It is seen that the distribution function approaches the Boltzmann form when
$T \approx \max(\nu_A, \nu_B)$.

\subsection{Approximation for all $T$}

Since the distribution of relative energies has a complicated form at finite temperatures,
an approximation might be helpful.
The distribution function of relative energies may be approximated, at all temperatures,
by a Fermi distribution 
\begin{equation}\label{F-Tappr}
 F_{T,\text{appr}}(\Erel) = \mathcal{N}_{\mu_\ast,T_\ast} \; \frac{1}{1+e^{(\Erel-\mu_\ast)/T_\ast}}
\end{equation} 
with the effective temperature $T_\ast$ from Eq.~\refe{Tast} and the effective chemical potential
\begin{equation}
 \mu_\ast = \frac{\nu_A+\nu_B}{2} = \frac{m_A \mu_B + m_B \mu_A}{m_A+m_B} .
\end{equation} 
This gives a reasonable approximation for the shape of $F_T(\Erel)$ at low temperatures.
For high temperatures, it approaches the exact distribution $F_\infty(\Erel)$ from Eq.~\refe{Boltz-3D}.

\section{Two dimensions}
\label{sec-2D}



Gases of ultracold atoms or molecules can be confined to thin layers.
For example, recent experiments at JILA~\cite{DeMiranda2011} investigated
a gas of KRb molecules in a pancake-shaped geometry
and measured the rate of chemical reactions;
related theoretical work 
considered ensemble-averaged rates 
(Boltzmann distribution in Ref.~\cite{Quemener2011}
and Fermi--Dirac distribution in Ref.~\cite{Pikovski2011})
for a two-dimensional geometry.
When the gas is confined to a thin layer,
one can describe collisions as if the particles were moving in a two-dimensional plane. 
The collisional cross-section, however, depends on the details of the
interparticle interaction and the confinement.
In this Section we will consider how the preceding results (Sec.~\ref{sec-rate} and Sec.~\ref{sec-DRE}) 
can be adapted to the case of two-dimensional scattering.



For two dimensions,
what was said at the beginning of Sec.~\ref{sec-rate} remains valid, but now 
in Eq.~\refe{rateeq} one should use the two-dimensional number densities
(dimensions $1/\text{length}^2$), and the two-dimensional scattering cross-section $\sigma'$ 
(dimensions $1/\text{length}$).
We will add a prime to two-dimensional quantities to avoid confusion.

The averaged two-dimensional collision rate $K'$, having dimensions $(\text{length})^2/\text{time}$, is 
\begin{equation}\label{K2d}
K' \!=\! \langle v \sigma'(p) \rangle \!=\! \iint \!\! v \sigma'(p) \bar{f}'_A(v_A) \bar{f}'_B(v_B) \, d^2v_A d^2v_B .
\end{equation} 
For velocity distributions which depend only on the energy of the particles,
changing to the center-of-mass (see Appendix~\ref{app-transf})
brings the integral to the following form:
\begin{equation}\label{K-F2d}
K' = \sqrt{\frac{2}{m}} \, \int_0^\infty  \!\! \Erel^{1/2} \, \sigma'(\Erel) F'(\Erel) \, d \Erel
\end{equation} 
with
\begin{equation}\label{F2d}
F'(\Erel) = \int_0^{2\pi} \! \frac{d\theta}{2\pi} \int_0^\infty \!\! f'_A (E_A) f'_B(E_B) \, d \Ecm .
\end{equation} 
Here $f'_{A,B}$ are the energy distribution functions, the
normalizations are $\int \! \bar{f}'(v) d^2v=1$ and $\int_0^\infty \! f'(E) dE=1$, so
$f'_{A,B}=(2\pi / m_{A,B}) \bar{f}'_{A,B}$.
The expressions for $E_A$ and $E_B$ in terms of $\Ecm$, $\Erel$ and $\theta$, which are needed
to evaluate the integral~\refe{F2d}, are given in Eq.~\refe{EAB-CR}.

The double integral in Eq.~\refe{F2d} can be transformed to the coordinate system $(x',y')$, see Appendix~\ref{app-transf},
which results in
\begin{equation}\label{F-xy}
 F'(\Erel) = \frac{1}{\pi \eta^2} \iint f'_A(E_A)f'_B(E_B) \, dx'dy' .
\end{equation} 
The integration runs over the whole space, $E_A$ and $E_B$ are expressed
through $x', y'$ using Eqs.~\refe{EAB-CR} and~\refe{Exy}, and $\eta$ is given in Eq.~\refe{eta}.

So far the discussion has been quite general, 
now we consider the case where the particles $A$, $B$ are fermions. 
At zero temperature the distribution functions for the particles $A$, $B$ are
\begin{equation}\label{f0-2D}
 f_0'(E) = \frac{1}{\mu}\, \theta(\mu-E) ,
\end{equation} 
with possibly different chemical potentials $\mu_A$, $\mu_B$.
The integral that results when \refe{f0-2D} is inserted in Eq.~\refe{F-xy}
has the geometric interpretation of the area of overlap of two circles.
For finite temperatures, the boundary of these circles becomes broadened by 
temperature, as before.

The area of intersection of two circles, with radii $r_1$ and $r_2$ separated by 
distance $d$ (see Fig.~\ref{fig:spheres}), is
\begin{multline}
 A_L \!=\! r_1^2 \arccos \Bigl(\frac{d^2+r_1^2 - r_2^2}{2dr_1}\Bigr) + r_2^2 \arccos \Bigl(\frac{d^2+r_2^2 - r_1^2}{2dr_2}\Bigr) \\
- \frac{1}{2} \sbsqrt{ (r_1^2 + r_2^2 +d^2)^2 - 2 (r_1^4 + r_2^4 + d^4) } .
\end{multline}
To include the cases where the circles are completely disjoint and where one circle is inside the other, we write
\begin{equation}\label{V2d}
 A(d;r_1,r_2) = 
\left\{
\begin{array}{ll}
0 & \quad\text{if } d^2 > (r_1+r_2)^2 \\[2pt]
\min(\pi r_1^2,\pi r_2^2)  & \quad\text{if } d^2 < (r_1 - r_2)^2 \\[3pt]
 A_L(d;r_1,r_2) & \quad\text{otherwise}
\end{array}
\right.
\end{equation} 

It follows that the distribution function of relative energies at $T=0$ is
\begin{equation}
 F'_0(\Erel) = \frac{2}{\pi \nu_A \nu_B} \; A(\sbsqrt{2\Erel}; \sbsqrt{\nu_A}, \sbsqrt{\nu_B} ) ,
\end{equation} 
where the $\nu_\alpha$ are given in Eq.~\refe{nu}.
This function has a similar shape as the 
corresponding function in three dimensions.

In the high-temperature case, the energies of the particles $A$, $B$ follow the Boltzmann distribution
\begin{equation}
f'_\infty(E)=\frac{1}{T} \, e^{-E/T}
\end{equation} 
with possibly different temperatures $T_A$, $T_B$.
Then the distribution of relative energies,
calculated from Eq.~\refe{F2d}, becomes a Boltzmann distribution in the relative energy
\begin{equation}
F'_\infty(\Erel) = \frac{1}{T_\ast} e^{-\Erel / T_\ast}
\end{equation} 
with the effective temperature $T_\ast$ as in Eq.~\refe{Tast}.
This result has the same form as in three dimensions.

\section{Threshold laws}
\label{sec-thres}


The dependence of the averaged collision rate
on the parameters of the two-species gas (mass, temperature, chemical potential)
can be found using the results of the preceding sections,
if the scattering cross-section is known. 
Some conclusions can be drawn, however, without
specific knowledge of the cross-section 
in the low-energy collisional regime where quantum threshold laws apply.
According to the Wigner threshold laws (see~\cite{Sadeghpour2000,Weiner1999} for a review), 
the cross-section at low energies scales with the relative energy as
\begin{equation}\label{thresh}
 \sigma(E) \propto E^p ,
\end{equation} 
where the exponent $p$ depends on the type of collision.
For elastic collisions with short-range forces in the \mbox{$\ell$-th} partial wave channel, 
we have $p=2\ell$. 
If molecules undergo chemical reactions upon collisions, the quantum threshold laws hold only
if the chemical reaction barrier is absent or much lower than the centrifugal barrier of 
scattering~\cite{Bell2009,Quemener2009}. 
For these exothermic barrierless reactions in the $\ell$-th partial wave, $p=\ell-1/2$.
In gases of ultracold atoms or molecules, one expects collisions to take place in the 
lowest partial wave channel that is allowed by exchange symmetry.

The averaged rate constant at zero temperatures for collisions obeying Eq.~\refe{thresh} 
can be obtained in closed form from Eqs.~\refe{K-F3d} and \refe{F-T0}. 
Here we only note the following property of this rate $K_0$. If the chemical potentials
are scaled as $\nu_A \to \lambda \nu_A$, $\nu_B \to \lambda \nu_B$, then $K_0 \to \lambda^{p+1/2} K_0$.
In particular, for $p=-1/2$ this means that $K_0$ is constant.

For finite temperatures, 
the exact dependence of the averaged rate constant on the 
reduced temperatures $\tau_A$, $\tau_B$ and chemical potentials
$\nu_A$, $\nu_B$ can be determined from the representation \refe{F-T}.
Inserting Eqs.~\refe{F-T} and~\refe{thresh} into Eq.~\refe{K-F3d} and performing the integration
over $\Erel$ first, one obtains expressions for $K$.
For $p=-1/2$, one finds again that $K$ is constant, 
and for other values of $p$ more complicated expressions result 
which involve combinations of Fermi--Dirac integrals
(see Appendix~\ref{app-finiteT} for their properties). 
In the strongly quantum degenerate regime ($\mu/T \gg 1$), 
one can obtain a low-temperature expansion applying Eq.~\refe{FD-expansion}.

A qualitative picture of the dependence of the averaged rate constant 
on the chemical potentials and temperature can be obtained by
approximating $F(\Erel)$ with a Fermi function with 
chemical potential $\mu_\ast$ and temperature $T_\ast$, cf.~Eq.~\refe{F-Tappr}.
Then, the averaged collision rate for cross-sections obeying Eq. \refe{thresh} has a simple form,
it scales for all temperatures as 
\begin{equation}\label{K-scaling}
 K_\text{appr} \propto T_\ast^{p+1/2} \frac{{\mathcal F}_{p+1}(\mu_\ast/T_\ast)}{{\mathcal F}_{1/2}(\mu_\ast/T_\ast)}
\end{equation} 
where the ${\mathcal F}$ are the Fermi-Dirac integrals.

For high temperatures, the distribution of relative energies approaches a Boltzmann distribution.
The threshold law \refe{thresh} then leads to the following well-known scaling of the collision rate
\begin{equation}
 K \propto T_\ast^{p+1/2} ,
\end{equation} 
which also follows from Eq.~\refe{K-scaling}.

\section{Conclusions}

Two-body collision or reaction rates in an ultracold gas are affected by quantum
statistics. The partial wave channel of the collision depends on the symmetry
of the particles involved; the density of final states may be modified 
(e.g.~Pauli blocking for fermions); and, finally, the available initial states depend on
the ensemble. In this paper, we have investigated initial-state effects for a gas
of two-species fermions with possibly different masses, particle densities, and temperatures.
Independently of the details of the collision, the ensemble-averaged collision rate is
the cross-section averaged with the distribution of relative energies.
This distribution function was calculated and its form was discussed in detail.
An approximation was proposed, which becomes exact in the high-temperature limit.
It was also shown how to transfer these results to particles confined to 
two dimensions. As an application of the results, we considered
the scaling of the averaged collision rates for the case where the cross-section follows
the Wigner threshold law.

The results will be of use for the analysis of experiments which measure collision or
reaction rates in gases of ultracold atoms or molecules. In particular, they will serve
to determine how the measured averaged rates in quantum degenerate two-species gases depend
on the temperature, the masses and densities of the particles.

I would like to acknowledge helpful discussions with L.~Santos, M.~Klawunn and C.~Salomon.

\appendix
\section{Transformation of coordinates}
\label{app-transf}

The coordinate transformation between the laboratory frame and the center-of-mass frame 
is given by the following expressions.
Two particles $A$, $B$ have masses $m_A$, $m_B$ and velocities $\vec{v}_A$, $\vec{v}_B$
in the laboratory frame. The velocities in the center-of-mass frame are
\begin{equation}\label{com1}
\vec{v}=\vec{v}_A-\vec{v}_B, \quad
(m_A+m_B)\vec{V}=m_A \vec{v}_A + m_B \vec{v}_B, 
\end{equation}
The corresponding momenta are
$\vec{p} = m (\vec{v}_A-\vec{v}_B)$ and
$\vec{P} = (m_A+m_B) \vec{V}$,
where 
$m=m_A m_B/(m_A+m_B)$. 
The related energies are
\begin{equation}\label{com3}
\begin{split}
& E_A=\frac{1}{2} m_A v_A^2, \qquad
E_B=\frac{1}{2} m_B v_B^2, \\[4pt]
& E_\text{cm}=\frac{1}{2}(m_A+m_B) V^2, \quad
E_\text{rel}=\frac{1}{2} m v^2.
\end{split}
\end{equation}
The angle between $\vec{v}$ and $\vec{V}$ is denoted by $\theta$.
The energies in the laboratory frame and center-of-mass frame are related by
\begin{equation}\label{EAB-CR}
\begin{split}
E_A &= \frac{m}{m_B} E_\text{cm} + \frac{m}{m_A} E_\text{rel} + \sqrt{ \frac{4 m E_\text{rel} E_\text{cm}}{m_A+m_B} } \cos \theta \\[2pt]
E_B &= \frac{m}{m_A} E_\text{cm} + \frac{m}{m_B} E_\text{rel} - \sqrt{ \frac{4 m E_\text{rel} E_\text{cm}}{m_A+m_B} } \cos \theta
\end{split}
\end{equation} 
and also
\begin{equation} 
E_\text{cm} + E_\text{rel} = E_A + E_B .
\end{equation} 
The Jacobian of the transformation from $(\vec{v}_A,\vec{v}_B)$ to $(\vec{v},\vec{V})$ is unity.

To discuss the integral over the center-of-mass energy $\Ecm$ and the angle $\theta$ in Eq. \refe{F3d}
(here $\Erel$ is a regarded as a constant parameter), we view $\theta$ as the azimuthal angle in
spherical coordinates, and introduce the coordinates $(x,y,z)$ as
\begin{equation}\label{Exyz}
\begin{split}
x &= \eta \sqrt{\Ecm} \cos \phi \sin \theta,
\quad
y = \eta \sqrt{\Ecm} \sin \phi \sin \theta, \\
z &= \eta \sqrt{\Ecm} \cos \theta,
\end{split}
\end{equation} 
with
\begin{equation}\label{eta}
 \eta = \frac{\sqrt{2 m_A m_B}}{m_A+m_B} .
\end{equation} 
The Jacobian is
\begin{equation}
 \frac{\partial(x,y,z)}{\partial(\Ecm,\theta,\phi)} = \frac{1}{2} \eta^3 \sin\theta \sqrt{\Ecm} .
\end{equation} 
The coordinates $(x,y,z)$ are proportional to the Cartesian components of the center-of-mass 
velocity $\vec{V}$ or the center-of-mass momentum $\vec{P}$.

For the two-dimensional integral over the center-of-mass energy, one can use the coordinates $(x',y')$
given by
\begin{equation}\label{Exy}
x'=\eta \sqrt{\Ecm} \cos \theta, \qquad
y'=\eta \sqrt{\Ecm} \sin \theta
\end{equation} 
with the Jacobian
\begin{equation}
 \frac{\partial(x',y')}{\partial(\Ecm,\theta)} = \frac{1}{2} \eta^2.
\end{equation} 

\section{Distribution function at finite temperature}
\label{app-finiteT}
The distribution function of the particles in a Fermi gas is
the normalized Fermi--Dirac distribution
\begin{equation}
 f_T(E) =  \mathcal{N}_{\mu,T} \, \frac{1}{1+e^{(E-\mu)/T}},
\end{equation} 
with the normalization constant 
\begin{equation}
\mathcal{N}_{\mu,T} = \frac{2}{\sqrt{\pi} T^{3/2} {\mathcal F}_{1/2}(\mu/T)},
\end{equation} 
such that $\int_0^\infty \! f(E) \sqrt{E} \, dE=1$.
Here 
\begin{equation}\label{FD-integral}
{\mathcal F}_j(x) = \frac{1}{\Gamma(1+j)} \int_0^\infty \!\!\frac{t^j }{1+e^{t-x}}dt  
\end{equation} 
is the Fermi-Dirac integral of order $j$~\cite{Kim-Blakemore}.
The expansion of this function for the strongly degenerate, low-temperature regime
($x \gg 1$) has the form
\begin{equation}\label{FD-expansion}
 {\mathcal F}_{j}(x) = \frac{x^{j+1}}{\Gamma(j+2)} 
\left[ 1 + C\, x^{-2} + \ldots \right]
\end{equation} 
for half-integer $j$, see Ref.~\cite{Kim-Blakemore} for more details.
For the non-degenerate, high-temperature regime ($x \ll -1$) we have ${\mathcal F}_j(x) = e^x$.

The chemical potential of a Fermi gas, for a fixed particle density, depends on the temperature. 
It is related to the Fermi energy $E_F=\mu(T \!=\! 0)$ by
\begin{equation}\label{mu-T}
 {\mathcal F}_{1/2}(\mu / T ) = \frac{4}{3\sqrt{\pi}} \Bigl(\frac{E_F}{T} \Bigr)^{3/2} .
\end{equation} 

In order to derive a representation for the integral \refe{F3d} when $f_A$ and $f_B$ are
Fermi-Dirac distributions, write
for the Fermi function $n_F(\xi) = (1+e^{\xi/T})^{-1} $ :
\begin{equation}\label{nF-rep}
n_F(\xi) 
=  -\int_\xi^\infty \!\! \frac{\partial n_F \!}{\partial \xi}(u) du 
=  -\int_{-\infty}^\infty \!\!\! \theta(u-\xi) \frac{\partial n_F \!}{\partial \xi}(u) du
\end{equation} 
Here 
\begin{equation}
 \frac{\partial n_F \!}{\partial \xi} = 
- \left[ 4T \cosh^2\left( \frac{\xi}{2T}\right) \right]^{-1}
\end{equation} 
Inserting Eq. \refe{nF-rep} into Eq. \refe{F3d} and changing the order of integration, we encounter
the $T=0$ distribution function of relative energies which was discussed in Sec~\ref{sec-DRE}.
The result is given in Eq. \refe{F-T}.


\begin{thebibliography}{99}

\bibitem{Ketterle2008}
W. Ketterle and M. W. Zwierlein,
{\em Making, probing and understanding ultracold Fermi gases},
in {\em Ultracold Fermi Gases}, 
Proceedings of the International School of Physics ``Enrico Fermi'' Varenna 2006, 
edited by M. Inguscio {\it et al} (IOS Press, Amsterdam, 2008).

\bibitem{Carr2009}
L. D. Carr {\it et al},
{\em Cold and ultracold molecules: science, technology and applications},
New J. Phys. {\bf 11}, 055049 (2009).

\bibitem{Bell2009}
M. T. Bell and T. P. Softley,
{\em Ultracold molecules and ultracold chemistry},
Mol. Phys. {\bf 107}, 99 (2009).

\bibitem{Weiner1999}
J. Weiner {\it et al},
{\em Experiments and theory in cold and ultracold collisions},
Rev. Mod. Phys. {\bf 71}, 1 (1999).

\bibitem{Krems2008}
R. V. Krems,
{\em Cold controlled chemistry},
Phys. Chem. Chem. Phys. {\bf 10}, 4079 (2008).

\bibitem{Julienne1989}
P. S. Julienne and F. H. Mies,
{\em Collisions of ultracold trapped atoms},
J. Opt. Soc. Am. B. {\bf 6}, 2257 (1989).

\bibitem{Ospelkaus2010}
S. Ospelkaus {\it et al},
{\em Quantum-state controlled chemical reactions of ultracold potassium-rubidium molecules},
Science {\bf 327} 853. (2019).

\bibitem{DeMiranda2011} 
M. H. G. de Miranda {\it et al},
{\em Controlling the quantum stereodynamics of ultracold bimolecular reactions},
Nature Physics {\bf 7}, 502 (2011).

\bibitem{Shuler1968}
K.~E.~Shuler, 
{\em Reaction Cross Sections, Rate Coefficients and Nonequilibrium Kinetics},
in {\em Chemische Elementarprozesse}, edited by H.~Hartmann (Springer, Berlin, 1968).

\bibitem{Sadeghpour2000}
H. R. Sadeghpour {\it et al},
{\em Collisions near threshold in atomic and molecular physics},
J. Phys. B: At. Mol. Opt. Phys. {\bf 33}, R93 (2000).

\bibitem{DeMarco1999}
B. DeMarco {\it et al},
{\em Measurement of p-Wave Threshold Law Using Evaporatively Cooled Fermionic Atoms},
Phys. Rev. Lett. {\bf 82}, 4208 (1999).

\bibitem{DeMarco2001}
B. DeMarco {\it et al},
{\em Pauli Blocking of Collisions in a Quantum Degenerate Atomic Fermi Gas},
Phys. Rev. Lett. {\bf 86}, 5409 (2001).

\bibitem{Napolitano1994}
R. Napolitano {\it et al}, 
{\em Line Shapes of High Resolution Photoassociation Spectra of Optically Cooled Atoms},
Phys. Rev. Lett. {\bf 73}, 1352 (1994).

\bibitem{Quemener2011}
G. Qu\'em\'ener and J. L. Bohn,
{\em Dynamics of ultracold molecules in confined geometry and electric field},
Phys. Rev. A {\bf 83}, 012705 (2011).

\bibitem{Quemener2009}
G. Qu\'em\'ener, N. Balakrishnan, and A. Dalgarno,
{\em Inelastic collisions and chemical reactions of molecules at ultracold temperatures},
in {\em Cold Molecules: Theory, Experiment, Applications}, edited by R.~Krems {\it et al} (CRC Press, 2009).


%

\bibitem{Kim-Blakemore}
J. S. Blakemore,
{\em Semiconductor statistics}
(Pergamon Press, Oxford, 1962);
R. Kim, M. Lundstrom,
{\em Notes on Fermi-Dirac Integrals},
arXiv:0811.0116


\bibitem{Pikovski2011}
A. Pikovski {\it et al},
{\em Nonlocal state swapping of polar molecules in bilayers},
Phys. Rev. A {\bf 84}, 061605(R) (2011).

\bibitem{Chevy-Radzhihovsky}
F. Chevy and C. Mora,
{\em Ultra-cold polarized Fermi gases},
Rep. Progr. Phys. {\bf 73}, 112401 (2010).
%
L. Radzihovsky and D. E. Sheehy,
{\em Imbalanced Feshbach-resonant Fermi gases},
Rep. Prog. Phys. {\bf 73}, 076501 (2010).


\bibitem{Partridge-Shin}
G. B. Partridge {\it et al}, 
{\em Pairing and Phase Separation in a Polarized Fermi Gas},
Science {\bf 311}, 503 (2006).
Y. Shin {\it et al},
{\em Phase diagram of a two-component Fermi gas with resonant interactions},
Nature {\bf 451}, 689 (2008).



\end{thebibliography}
\end{document}